\documentclass[
notitlepage, 
reprint,twocolumn, pra,
amsmath,amssymb,superscriptaddress,nofootinbib,showpacs,aps
]{revtex4-1}

\pdfoutput=1
\usepackage{graphicx}
\usepackage{bm}
\usepackage{times}
\usepackage{comment}
\usepackage{dsfont}
\usepackage{color}
\usepackage{float}
\usepackage{soul}
\usepackage[dvipsnames]{xcolor}

\definecolor{blue}{RGB}{0,0,255}
\definecolor{teal}{RGB}{0,100,100}

\newcommand{\MC}{\textcolor{black}}
\usepackage[colorlinks,citecolor=blue,linkcolor=blue,urlcolor=blue]{hyperref}

\graphicspath{{"fig/"}}

\def\BE{\begin{equation}}
\def\EE{\end{equation}}
\def\BY{\begin{eqnarray}}
\def\EY{\end{eqnarray}}
\def\BI{\begin{itemize}}
\def\EI{\end{itemize}}
\def\L{\label}
\def\nn{\nonumber}
\def\({\left (}
\def\){\right)}
\def\[{\left [}
\def\]{\right]}
\def\<{\langle}
\def\>{\rangle}

\def\BA{\begin{array}}
\def\EA{\end{array}}

\def\dd{\delta}
\def\D{\Delta}
\def\w{\omega}

\def\G{\Gamma}

\def\s{\sigma}
\def\+{\dag}
\def\8{\infty}

\def\l{\lambda}
\def\ii{\text{i}}
\def\={\approx}

\def\->{\rightarrow}

\def\qq{\textbf{q}}

\def\rr{\boldsymbol{\rho}}
\newcommand{\ud}{\,\mathrm{d}} 

\def\ts{\text{s}}

\def\tp{\text{p}}

\def\um{$\rm{\mu m}$}
\def\um1{$\rm{\mu m}^{-1}$}

\begin{document}
\title{
Reconstructing 2D spatial modes for classical and quantum light
}

\author{Valentin A. Averchenko}
\email{valentin.averchenko@gmail.com}
\affiliation{St. Petersburg State University, Ul'yanovskaya street 3, 198504 Saint Petersburg, Russia}
\author{Gaetano Frascella}
\affiliation{Max Planck Institute for the Science of Light, Staudtstr. 2, 91058
Erlangen, Germany.}
\affiliation{University of Erlangen-Nuremberg, Staudtstr. 7/B2, 91058 Erlangen,
Germany.}
\author{Mahmoud Kalash}
\affiliation{Max Planck Institute for the Science of Light, Staudtstr. 2, 91058
Erlangen, Germany.}
\affiliation{University of Erlangen-Nuremberg, Staudtstr. 7/B2, 91058 Erlangen,
Germany.}
\author{Andrea Cavanna}
\affiliation{Max Planck Institute for the Science of Light, Staudtstr. 2, 91058
Erlangen, Germany.}
\author{Maria V. Chekhova}
\affiliation{Max Planck Institute for the Science of Light, Staudtstr. 2, 91058
Erlangen, Germany.}
\affiliation{University of Erlangen-Nuremberg, Staudtstr. 7/B2, 91058 Erlangen,
Germany.}

\date{\today}

\begin{abstract}

\MC{We propose a method for finding 2D spatial modes of thermal field through a direct measurement of the field intensity and an offline analysis of its spatial fluctuations.}
%
%
Using this method, \MC{in a simple and efficient way we reconstruct the modes of a multimode fiber and the spatial Schmidt modes of squeezed vacuum generated via high-gain parametric down conversion. The reconstructed shapes agree with the theoretical results.}  

\end{abstract}
%

\maketitle
\section{Introduction}

One of the main tasks of statistical optics is to determine the coherent properties of an electromagnetic field and, as a consequence, unveil information about the generation and propagation of the radiation.
An important concept is \MC{the one of radiation modes, i.e., solutions to the wave equation. Modes can be viewed as space/time field distributions (or, alternatively, field distributions in wavevector/frequency space) where the field is coherent with itself but incoherent with the field in other modes. Modes are most commonly chosen as plane monochromatic waves, but there are more elegant ways to define them. Examples, further used in this paper, are coherent modes of thermal light~\cite{Wolf:82, MandelWolf}, Schmidt modes of a bipartite quantum system~\cite{Law2000}, and the spatial modes of a multimode fiber. In all three cases, the retrieval of mode shapes is crucial but not always a simple task.}

In this work, we \MC{propose a simple method to retrieve the spatial modes of multimode radiation. Using this method, we solve two important practical problems from classical and quantum optics. Namely, we experimentally reconstruct two-dimensional (2D) spatial modes of a multimode optical fiber and the Schmidt modes of the quantum radiation generated through the high-gain parametric down-conversion (PDC).}

The experimental reconstruction of the spatial eigenmodes of a fiber, especially \MC{a microstructured one}, is crucial since the actual modes can deviate from the simulated ones. Some of the reconstruction methods face computational complexity~\cite{Shapira:05} and \MC{require} sensitive alignment of interferometers~\cite{Lyu:17} or cavities~\cite{Ahn:05}. The most established technique is the $S^2$-imaging~\cite{Nicholson:08}, \MC{which} relies on the interference occurring inside the fiber between the fundamental Gaussian mode and the higher-order modes, therefore it is alignment-free. Yet the wavelength of the coupled light needs to be scanned with a tunable source and the analysis can be time-consuming for the interference patterns, originally measured with a space-scanning fiber tip~\cite{Nicholson:08}, and more recently with a camera~\cite{Nguyen:12}.

\MC{For PDC radiation, the multimode structure is both an advantage, because it provides an additional resource in quantum communication~\cite{Zhou:16} and sensing~\cite{Kutas:20}, and a challenge to describe. The use of the Schmidt modes framework simplifies the photon correlations~\cite{Law2000} for both low-gain~\cite{miatto_cartesian_2012} and high-gain PDC~\cite{Sharapova:15}: a single Schmidt mode has photon-number correlations only with itself or with a single matching mode. Experimentally finding the Schmidt-mode profiles is therefore important, but difficult to do in 2D space because the standard procedure for doing this, \textit{singular-value decomposition}, is only defined for one dimension. Until now, 2D coherent modes of PDC have never been reconstructed, although four-dimensional (4D) joint probability distributions for PDC have been measured~\cite{Reichert:18}. Instead, because higher-order spatial modes are required for quantum communication, several groups reconstruct the modes of the PDC radiation only in the azimuthal degree of freedom, i.e. the orbital angular momentum spectrum~\cite{Mair:01,Pires:10,Kulkarni:17}. Alternatively, Schmidt modes can be reconstructed in vertical and horizontal Cartesian dimensions separately, if there is a corresponding symmetry~\cite{Straupe:11}. However, this is not always the case.}

\MC{Our approach to reconstruct 2D spatial modes includes the following steps.
First, we directly measure} 
\footnote{
The measurement of spatial intensity correlations can be done with a standard camera.
Measuring spatial field correlations requires measurement of the field interference.
}
the intensity correlation function, which allows \MC{us to find the field correlation function for light with thermal statistics. 
This is indeed the case for the output PDC radiation provided that only signal or idler radiation is measured, and also for a fiber fed with pseudo-thermal light.}  

The \MC{calculated spatial field correlation functions form 4D arrays. We then
convert each 4D array into a 2D one} using an array flattening procedure.
\MC{Finally, a standard diagonalization of the resulting array yields} 2D profiles of the coherent modes of the field and their integral intensities.

In the case of PDC, the coherent modes found with this procedure coincide with the Schmidt modes of the down-converted radiation~\cite{Felix}.

Our method, applicable to a large number of cases, consists of very simple measurements and data elaboration and promises to outperform all other methods.

Further, we describe the theory of our method in Section~\ref{sec:th}. Subsection~\ref{sec:CFs} defines the field and intensity correlation functions and Subsection~\ref{sec:Siegert} provides the link between them for thermal light. The procedure of converting a 4D array into a 2D one is described in Subsection ~\ref{sec:representation}, and the coherent-mode representation, in Subsection~\ref{sec:coh}. Section~\ref{sec:exp} is devoted to the experiment: subsection~\ref{sec:highgainPDC} deals with the mode reconstruction for the high-gain PDC and Subsection~\ref{sec:MMF}, for the multimode fiber. Сonclusions are made in Section~\ref{sec:concl}. 
In the Appendix~\ref{app:modes_theory} an analytical model for calculating the field correlation function of twin beams of the high-gain PDC is presented.

\section{Theory\label{sec:th}}
In this section, we review the fundamental quantities of statistical optics like the field and intensity correlation functions (CFs) and their interconnection for light with thermal statistics.
Then we propose a method to reduce the dimensionality of such quantities by re-organization of the distributions. Finally, we discuss the importance of the coherent modes representation for the field CF. 

 \subsection{Field and intensity correlation functions\label{sec:CFs}}

	\begin{figure}[b]
	\center{\includegraphics[width=.5\linewidth]{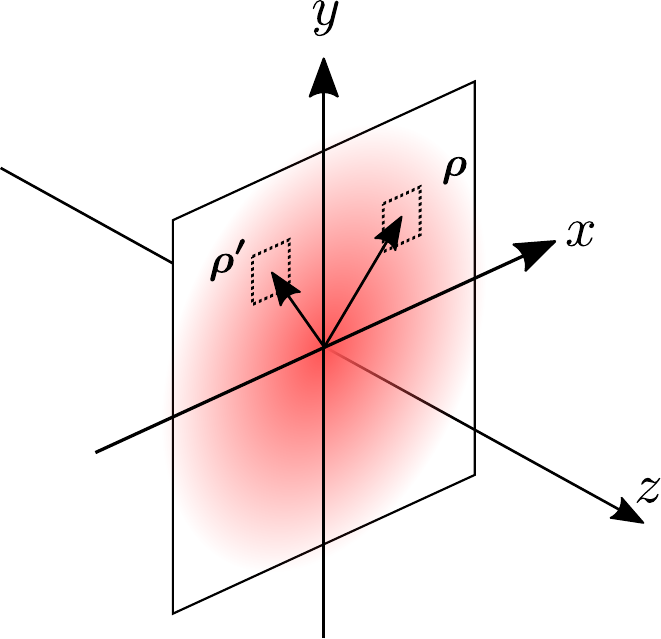}}
	\caption{Cross section of a light beam and chosen position vectors $\rr$ and $\rr'$.}
	\label{fig:model}
	\end{figure}

Consider a monochromatic beam of light with a frequency $\w_0$ and fixed polarization propagating along the $z$-axis (Fig.~\ref{fig:model}).
The electric field of the beam at a given cross-section with coordinates $\rr = (x,y)$ and at a time $t$ can be modelled as
	\begin{align}
	E(\rr, t) \propto a(\rr) e^{-i \w_0 t} + a^\+(\rr) e^{i \w_0 t},
	\L{def:a}
	\end{align}
where $a(\rr)$ stands for the complex amplitude of the field at a point $\rr$.
In the quantum theory $a(\rr)$ is an operator, with the commutation relation $[a(\rr), a^\+(\rr)] = \delta(\rr-\rr')$ (for example, see \cite{Kolobov1999}).
The proportionality coefficient in the expression (\ref{def:a}) is chosen in such a way that the quantity
	\begin{align}
	I(\rr) = a^\+(\rr) a(\rr)
	\end{align}
gives the  photon-flux density in  photons  per  unit area of the beam cross-section.

The spatial CF of the field amplitude reads
	\begin{align}
	G^{(1)}(\rr,\rr') \equiv \<a^\+(\rr) a(\rr')\> ,
	\L{eq:G1_def}
	\end{align}
where brackets stand for classical/quantum ensemble averaging.
The function represents correlations of the field at a pair of points/pixels in the transverse plane (Fig.~\ref{fig:model}).
Note that the mean intensity at point $\rho$ is given by the diagonal value of the CF:
	\begin{align}
	\<I(\rr)\> = G^{(1)}(\rr,\rr).
	\L{I_def}
	\end{align}
If the function is known in one transverse plane, then it can be calculated in another transverse plane along the light propagation using the corresponding propagation equation for the CF, for example, from \cite{Goodman}.

There are several experimental methods to reconstruct the field CF (for example, see \cite{Raymer1994} and references therein).
In this work, we reconstruct the first-order CF of the field from the measurement of the intensity CF, which reads
	\begin{align}
	\<I(\rr) I(\rr')\> 	&= \<a^+(\rr) a(\rr) a^+(\rr') a(\rr')\>.
    \L{eq:G2_def}	
	\end{align}

It characterizes correlations of intensities at two points of the beam cross-section.
The function can be reconstructed via repetitive measurements of the point-by-point cross-section intensity and calculating pairwise correlations of intensities (e.g. using a camera and processing the data).

\subsection{Link between first- and second-order correlation functions for thermal light\label{sec:Siegert}}

In this paper, we consider light with \MC{thermal statistics. Such light is emitted by thermal/chaotic sources, for example, through the} spontaneous uncorrelated emission of many atoms. 
Also, the signal/idler beams generated by PDC have thermal statistics. 
\MC{For thermal light, field and intensity CFs (\ref{eq:G1_def},\ref{eq:G2_def}) are related as}~\cite{loudon_quantum_2000}, 
	\begin{align}
	     \<I(\rr) I(\rr')\>=& |G^{(1)}(\rr,\rr')|^2 \nn\\
	     &+ \<I(\rr)\> \<I(\rr')\> + \<I(\rr)\> \dd(\rr-\rr').
	\label{eq:G2fromG1}
	\end{align}
The last term is due to the quantization of the field energy. Formally, it appears \MC{after the} normal ordering of the operators in the expression (\ref{eq:G2_def}), using \MC{commutation relations. This term, known as the shot noise, is independent of the light statistics~\cite{Kolobov1999} and describes} uncorrelated field intensities at different spatial points of the transverse plane.
For intense enough light (number of photons per coherence area is much greater than one), its contribution is relatively small and can be neglected. \MC{In this case, Eq.~(\ref{eq:G2fromG1}) is} known as the Siegert relation \cite{MandelWolf}.

This relationship between CFs has been used in Hanbury Brown and Twiss (HBT) experiment to solve the inverse problem: reconstruct the field amplitude correlations and  estimate the characteristics of the emitters, like stars, by measuring the intensity correlations.
Indeed, one can estimate the modulus of the first-order CF by inverting Eq.~(\ref{eq:G2fromG1}):
	\begin{equation}
	\begin{array}{c}
	|\tilde G^{(1)}(\rr,\rr')| \approx \sqrt{\<I(\rr) I(\rr')\> - \<I(\rr)\>\<I(\rr')\>},
	\end{array}\L{G1fromG2}
	\end{equation}
where the shot-noise term is neglected.
The right-hand side of Eq.~(\ref{G1fromG2}) is the square root of the intensity covariance, which characterizes the correlation of the intensity fluctuations at two points of the beam cross-section:
	\begin{align}
	& \text{Cov}(\rr,\rr') = \<\dd I(\rr) \dd I(\rr')\>,
	\end{align}
where $\dd I(\rr) = I(\rr) - \<I(\rr)\>$.
\MC{Equation~(\ref{G1fromG2}) allows one to restore the first-order CF of a thermal field completely provided there is no phase modulation,} i.e. $G^{(1)} = |G^{(1)}|$.

\subsection{\MC{Full dimensionality of the correlation functions}} \label{sec:representation}

In this section, we consider the full dimensionality of only the first-order field CF (\ref{eq:G1_def}), but all the following statements can be easily applied to the intensity CF (\ref{eq:G2_def}). The function $G^{(1)}(\rr,\rr')$, containing information on the \MC{correlations of the} complex field amplitude for all pairs of points, depends in general on four spatial scalar coordinates. Indeed, the position of each point can be specified by two Cartesian coordinates $\rr=(x,y)$, as shown in Fig.~\ref{fig:model}, or by two polar coordinates $(\rho,\phi)$.
The values of the functions can be arranged in 4-dimensional arrays~\footnote{We consider the values of functions for a discrete set of points because in the experiment we use a camera with pixels.} but one cannot simply visualize such arrays and analyze correlations in this representation.

Here are some examples of situations where this problem does not occur. First, when the radiation field is statistically homogeneous and isotropic in the cross-section, the CF depends only on the distance between the cross-sectional points, namely on a scalar: $G^{(1)}(\rr,\rr') = G^{(1)}(|\rr-\rr'|)$.
Second, when the field properties are symmetrical with respect to the propagation axis, the CFs are factorable in two variables, e.g. in $\rho$, $\phi$ and $G^{(1)}(\rr,\rr') = G^{(1)}_R(\rho,\rho')\,G^{(1)}_A(\phi,\phi')$. In this particular case, radial and azimuthal CFs, respectively indicated with subscripts $R$ and $A$, can be analyzed independently and the values of each function $G^{(1)}_{R,A}$ can be also arranged in 2-dimensional arrays and analyzed (see, for example, \cite{Frascella:19}).
\MC{Similarly, in some cases the factorization takes place for $x$ and $y$ Cartesian coordinates~\cite{Straupe:11}}.

In the general case of spatially non-uniform and non-isotropic field in the transverse plane, the visualization problem of field correlations and their analysis can be treated as follows.

\begin{figure}[b]
	\center{\includegraphics[width=.5\linewidth]{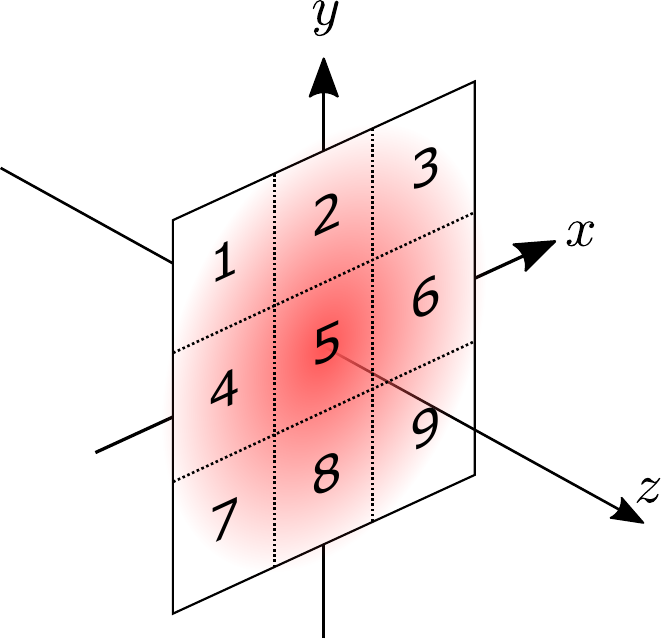}}
	\caption{Dividing the cross section into 9 pixels and numbering them in a row-major order.}
	\label{fig:numbering}
	\end{figure}
We re-organize the 2D field distribution by re-numbering all the transverse-plane points in the order shown in Fig.~\ref{fig:numbering}. Of course, different ways of re-organization into 1D arrays are possible. In the next step, we arrange the 4D distribution of the CF into a 2D array as follows:
\begin{align}
\{G^{(1)}(\rr_{n},\rr_{m})\} \rightarrow 
 \begin{pmatrix}
  G^{(1)}(\rr_1,\rr_1) & G^{(1)}(\rr_1,\rr_2) & \cdots \\
  G^{(1)}(\rr_2,\rr_1) & G^{(1)}(\rr_2,\rr_2) & \cdots \\
  \vdots  & \vdots  & \ddots 
 \end{pmatrix}.
 \L{G1_2D}
\end{align}
Here the first line of the array contains information on the \MC{correlations of the field at the first point with the field at the first point (auto-correlation), at} the second point and so on.
The second line contains information on the correlations of the field at the second point with the field at the first point, at the second point etc., while the $n$-th line of the array contains information about the field correlations at the $n$-th point with the fields at all other points.
This reorganization procedure leads to a visual 2D representation of the field correlation data at all points of the transverse plane.

The above procedure to \MC{replace a 4D array with a 2D} one is similar to tensor reshaping, i.e. a bijective map between an order-$d$ tensor and an order-$k$ tensor, where $k<d$.
Particular examples of the reshaping are called array/tensor flattening, matricizations, unfolding.

Besides the visual representation of \MC{2D field correlations, such an approach allows us to calculate numerically the} coherent modes of the field, which are discussed in the next section.


\subsection{Coherent modes\label{sec:coh}}

The first-order CF of the field is Hermitian, i.e. $G^{(1)}(\rr,\rr') = \left(G^{(1)}(\rr',\rr)\right)^*$, according to its definition~(\ref{eq:G1_def}).
Thus, according to Mercer's theorem, it admits the representation
	\begin{align}
	& G^{(1)}(\rr,\rr') = \sum_m \l_m u_m^*(\rr) u_m(\rr'),
	\L{G1decomp}
	\end{align}
whose eigenvalues $\l_m$ and eigenfunctions $u_m(\rr)$ satisfy the integral equations
	\begin{align}
	& \int G^{(1)}(\rr,\rr') u_m(\rr') \ud \rr' = \l_m u_m(\rr).
	\L{eigen_problem}
	\end{align}
If the eigenvalues are not degenerate, they can be ordered, for example, in descending order and numbered accordingly.
Also, they can be numbered according to the spatial characteristics of the eigenfunctions. In this case, multiple indices can be used. 
The values are non-negative, and the functions are orthogonal and typically taken to be orthonormal. \MC{Equation~(\ref{G1decomp}) is called the coherent-mode representation of the first-order CF and the functions $u_m(\rr)$, the} spatial coherent modes of the field~\cite{Wolf:82, MandelWolf}.

Using Eq.~(\ref{I_def}), one can see that the average beam intensity at a given cross-sectional point $\rr$ is a sum of modulus-squared coherent modes multiplied by the weights $\l_m$: 
	\begin{align}
	\label{eq:intensity}
	\<I(\rr)\>= \sum_m \l_m |u_m(\rr)|^2.
	\end{align}
Thus, $\l_m$ can be considered as the integral intensity of the coherent mode with index $m$.
The effective number of coherent modes can be estimated with
\footnote{
One notes that $\sum \lambda_m = 1 $ when the eigenvalues a normalized.
}
	\begin{align}
	\label{eq:numbermodes}
	& {\cal K} = \frac{(\sum \l_m)^2}{\sum \l_m^2}.
	\end{align}
If there is just a single term in the decomposition in Eq.~(\ref{G1decomp}), then ${\cal K}=1$ and the beam is referred to as single-mode. Then the first-order CF is factorable and the light field is fully spatially coherent.

The field representation as a sum of fields of coherent modes has several applications \cite{gbur_structure_2010}. We can stress few of them.
First, Eq.~(\ref{G1decomp}) shows that coherent-mode representation gives information about the field correlations in a cross-section: knowing the coherent modes, one can restore the first-order CF. 
This is convenient for describing spatial correlations in the general case of non-uniform and non-isotropic fields, where the values of the CF form a multidimensional array.
Indeed, if a CF is represented by a 4D array, then coherent modes constitute 2D arrays and can be visualized in a 2D density plot. 
Second, the profiles of coherent modes and the distribution of their integral intensities give an additional insight into the light generation process \cite{Roman2014, De:19}.
Third, the propagation of a partially-coherent light beam can be viewed as an independent propagation of fully coherent modes.
Fourth, the representation of \MC{a thermal field as a sum of coherent-mode fields enables solving} a number of problems, e.g. spatial  filtering with minimal losses to obtain fully coherent radiation. Indeed, from Eq.~(\ref{G1decomp}) it follows that it is necessary to filter all modes, except the mode with the highest eigenvalue.

\MC{The coherent-mode decomposition of a 2D CF can be done as follows.}
First, the CF values for the discrete point set are presented
\footnote{
An alternative numerical approach - representation of values of a multidimensional correlation function in a discrete basis set \cite{Flewett:09, Annamalai:11}}
as a \MC{Hermitian 2D matrix as described by Eq.~(\ref{G1_2D}).
Then, the search for the matrix's} eigenvalues and eigenvectors is performed.
Each eigenvector is then transformed into a matrix as
\begin{align}
\begin{pmatrix}
  u_m(\rr_1)\\
  u_m(\rr_2)\\
  \vdots
\end{pmatrix}
\overset{\rr = (x,y)}{\rightarrow}
\begin{pmatrix}
  u_m(x_1,y_1) & u_m(x_2,y_1) & \cdots \\
  u_m(x_2,y_1) & u_m(x_2,y_2) & \cdots \\
  \vdots  & \vdots  & \ddots 
 \end{pmatrix}.
 \L{vector-to-matrix}
\end{align}
This procedure is inverse to the unfolding procedure.
As a result, one gets 2D "profiles" of coherent modes of the field with a given first-order CF $G^{(1)}$.

\section{Experiment\label{sec:exp}}

Here we consider two sources of light: one of the twin beams \MC{generated through high-gain PDC and the output radiation of a multimode fiber fed} with thermal light.
\MC{Both sources, as mentioned in the Introduction, have thermal statistics.}
\begin{figure}
	\center{\includegraphics[width=.75\linewidth]{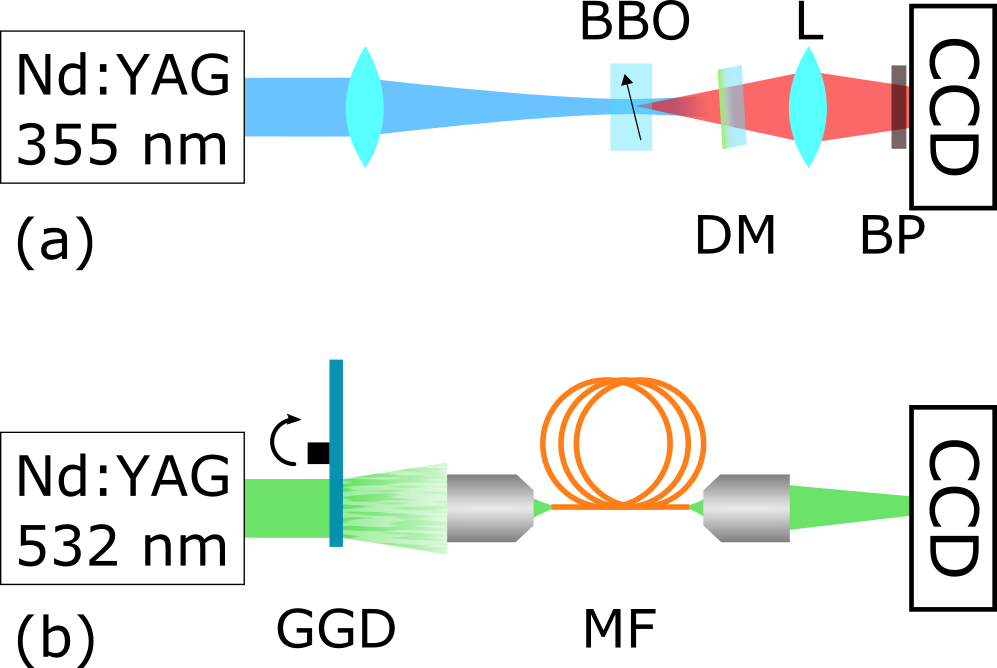}}
	\caption{Experimental setup for the reconstruction of the modes for high-gain PDC (a) and for a multimode fiber (b).}
	\label{fig:setup}
	\end{figure}
We measure the spatial distribution of the intensity fluctuations in the beam cross-section for each source with the experimental setups shown in Fig.~\ref{fig:setup} and reconstruct the first-order CFs and the coherent modes of the fields.
The obtained experimental results \MC{we compare with the ones of the theoretical models.}

\subsection{High-gain PDC}
\label{sec:highgainPDC}

In the first case considered, the light is generated via high-gain PDC in a second-order nonlinear transparent crystal, as shown  Fig.~\ref{fig:setup}a. Pump photons can be annihilated to create twin beams, usually referred to as signal and idler and distinguished by polarization, frequency, or propagation direction. 
Conservation of momentum during the process leads to \MC{quantum photon-number correlations between groups of signal and idler wavevector (plane-wave) modes}
\footnote{\MC{There are also photon-number correlations between frequency modes. However, here we focus on the wavevectors} and do not consider the frequency degree of freedom.}.
However, there exists a basis of signal and idler spatial modes, so-called Schmidt modes of PDC, \MC{in which the correlations are simplified:
each signal mode is only correlated in photon number} with a single matching idler mode~\cite{Law:04}.
The knowledge of the Schmidt modes is important for quantum information applications.

One can show that \MC{the Schmidt modes of the bipartite system formed by both signal and idler beams coincide with the coherent modes of the two subsystems taken separately~\cite{Felix}. Because each of the twin beams has thermal statistics, based on the results of Section~\ref{sec:th}, the Schmidt modes can be reconstructed from the analysis of the intensity CF of just one beam.}

To generate twin beams through type-I collinear degenerate PDC, we use a $2$ mm $\beta$-barium borate (BBO) crystal and the pump at $354.67$ nm from the third-harmonic beam of a Nd:YAG laser. The $18$ ps pulses at a repetition rate $1$ kHz and average power $117$ mW are needed to reach the high-gain regime. After the generation of PDC radiation, the pump is rejected with a dichroic mirror (DM).

\begin{figure}[H]
	\center{\includegraphics[height=.47\linewidth]{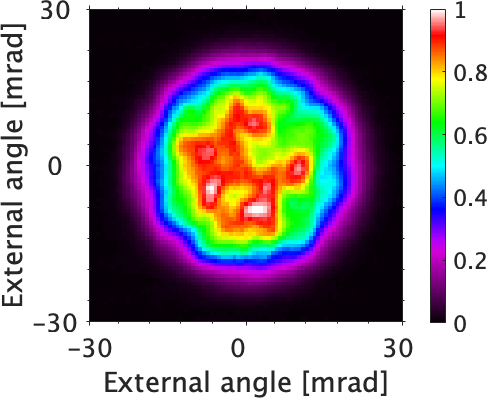}}
	\caption{Single-shot intensity distribution for the PDC emission in the far field as a function of the external angles.}
	\label{fig:shotPDC}
	\end{figure}

A set of $3000$ single-shot intensity distributions is captured with a charge-couple device (CCD) camera in the focal plane of lens L (focal length $f=40$ mm). Fig.~\ref{fig:shotPDC} shows the far-field intensity distribution, plotted versus two Cartesian angles. These angles -- called external because computed outside the crystal -- are found as the ratio of the Cartesian transverse wavevector components and the signal wavevector modulus. 

We use a bandpass filter (BP) centered at $700$ nm with a bandwidth of $10$ nm attached to the camera for frequency filtering. By using a central wavelength detuned from the degenerate one ($709.3$ nm), we remove the idler modes matching with the signal. In this way, we select only one of the twin beams and the typical cross-correlation of intensity fluctuations between signal and idler modes disappears~\cite{Beltran:17,Frascella:19}. Since the detuning from degenerate wavelength is small, the reconstructed modes do not differ from the eigenmodes of the degenerate PDC.

Given the high number of frequency modes selected, the intrinsic fluctuations of the twin-beam power are weak. But the pump excess noise induces additional power fluctuations and, in this case, we find that the first coherent mode erroneously resembles the average intensity distribution. To avoid this, we normalize each spectrum to the integral intensity and thus eliminate the effect of all pulse-to-pulse power fluctuations. This normalization results in a small negativity of the covariance distribution, which, as an artifact of this procedure, is rejected by taking the real part of the square root in Eq.~(\ref{G1fromG2}).

\begin{figure}[H]
	\center{\includegraphics[height=.47\linewidth]{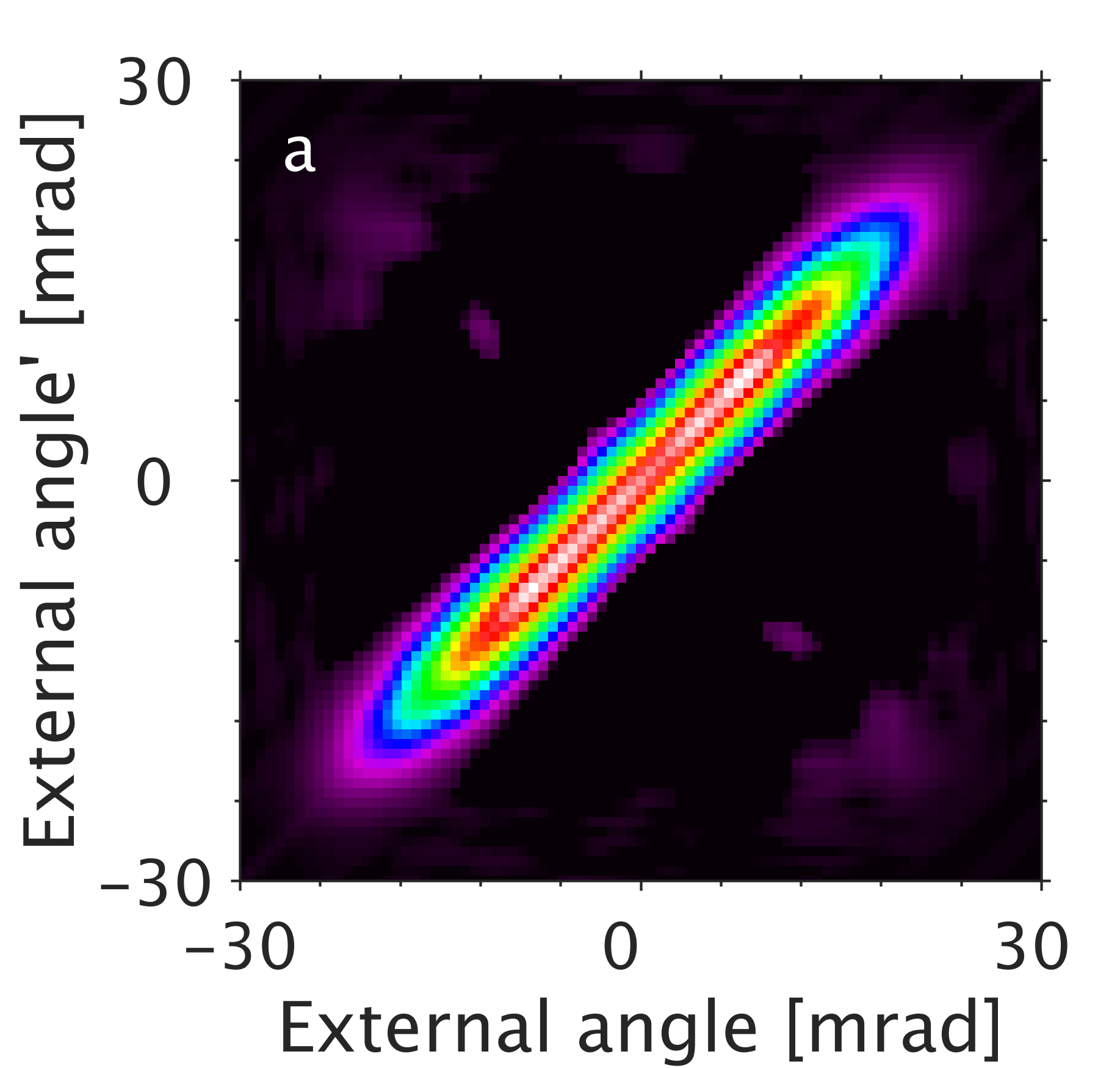}\includegraphics[height=.47\linewidth]{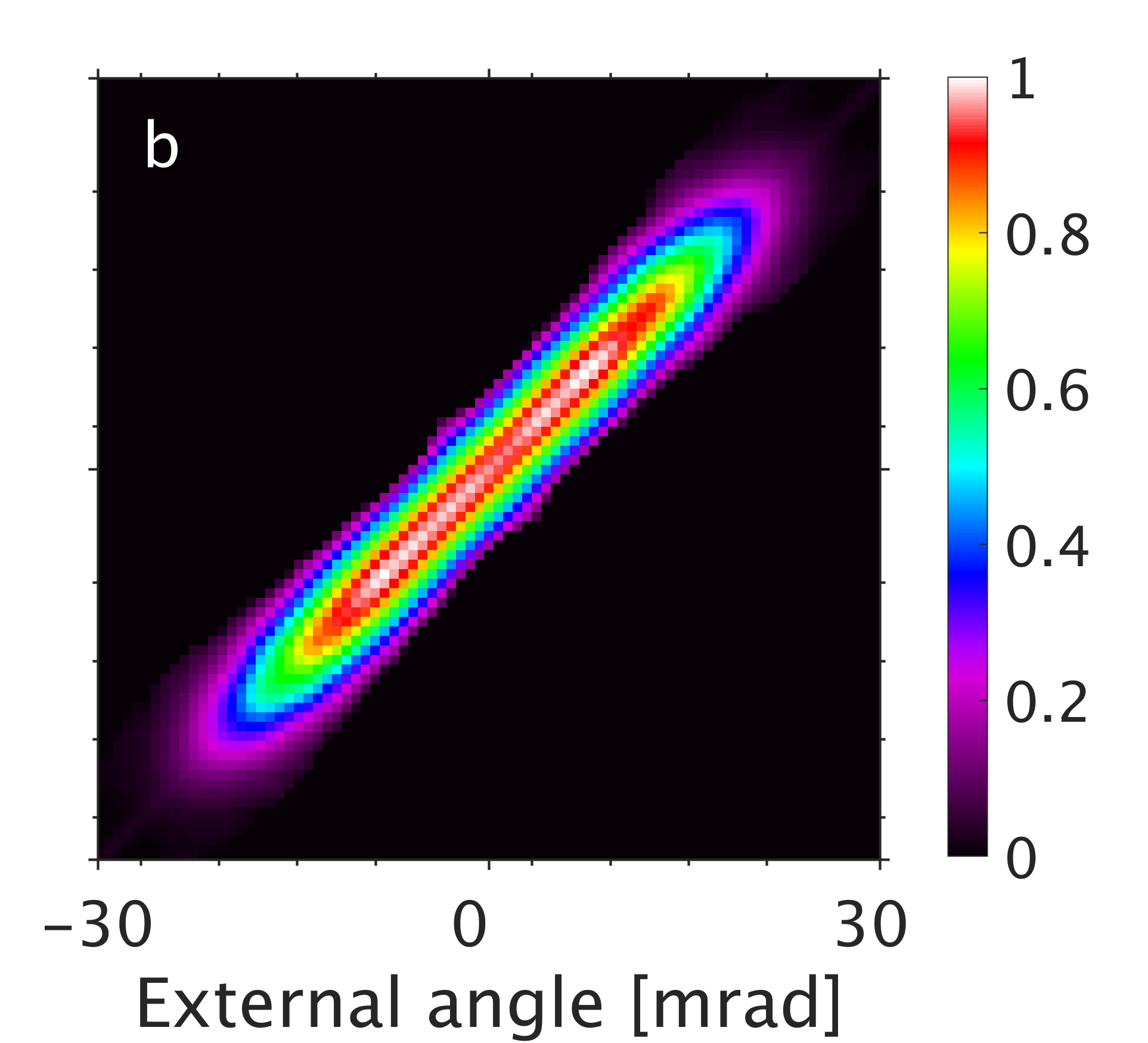}}
	\caption{$|G^{(1)}|$ distribution for the PDC emission with vertical coordinate fixed to zero as a function of the external angles with (a) and without noise (b).}
	\label{fig:covPDC}
	\end{figure}

The far-field intensity correlation distribution from the reorganization procedure explained in Sec.~\ref{sec:representation} would require a high-resolution plot. Therefore, we show only a distinctive feature of the correlations from a 1D cut of Fig.~\ref{fig:shotPDC}, namely with the vertical coordinate fixed to zero. Following the prescription in Eq.~(\ref{eq:G2fromG1}), from the covariance distribution we obtain the $|G^{(1)}|$ distribution shown in Fig.~\ref{fig:covPDC} (a). Here, one can see correlations for the external angles equal within a $\sim7$ mrad range. We remove the noise present in the distribution -- see Fig.~\ref{fig:covPDC} (b) -- to obtain better results in the reconstruction.

    \begin{figure}[H]
	\center{\includegraphics[width=\columnwidth]{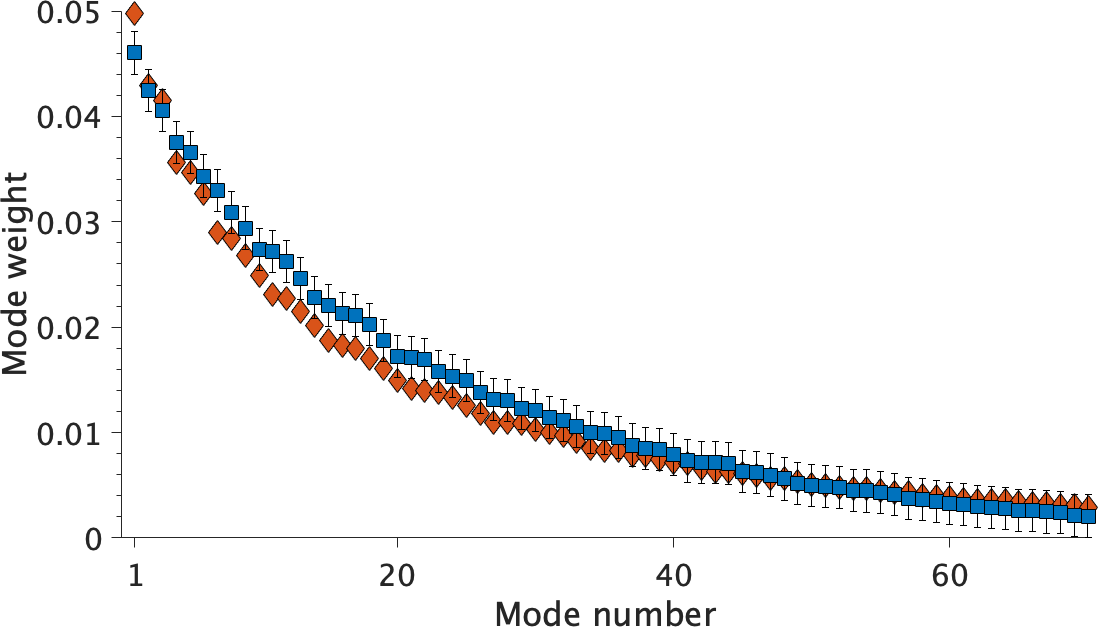}}
	\caption{Weights of the Schmidt modes for PDC light reconstructed from the experiment (blue squares) and from the simulation (red diamonds). The weights are normalized to their sum up to the 200th mode.}
	\label{fig:weights_exp}
	\end{figure}

	\begin{figure*}[p]
	\center{\includegraphics[width=.85\linewidth]{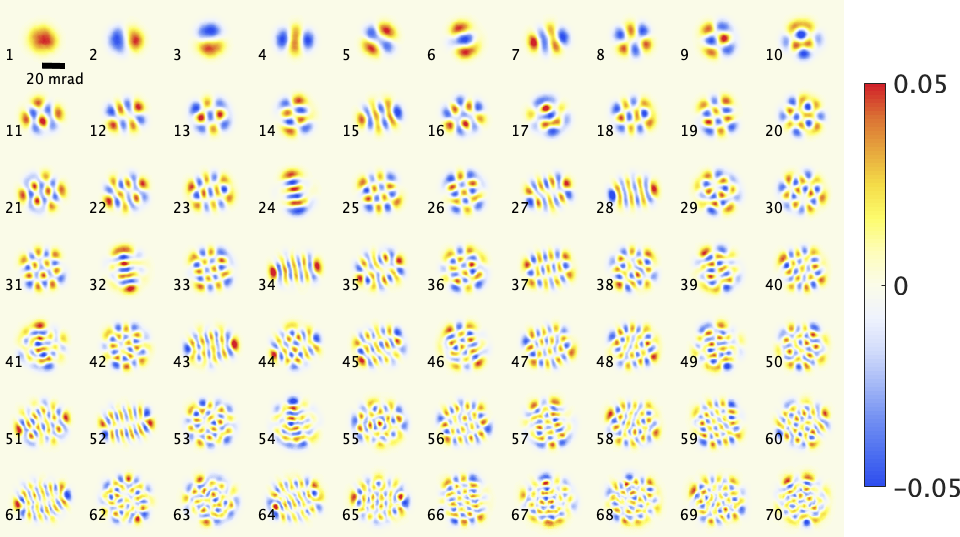}}
	\caption{First 70 experimentally reconstructed modes for PDC radiation.}
	\label{fig:modes_exp}

	\center{\includegraphics[width=.85\linewidth]{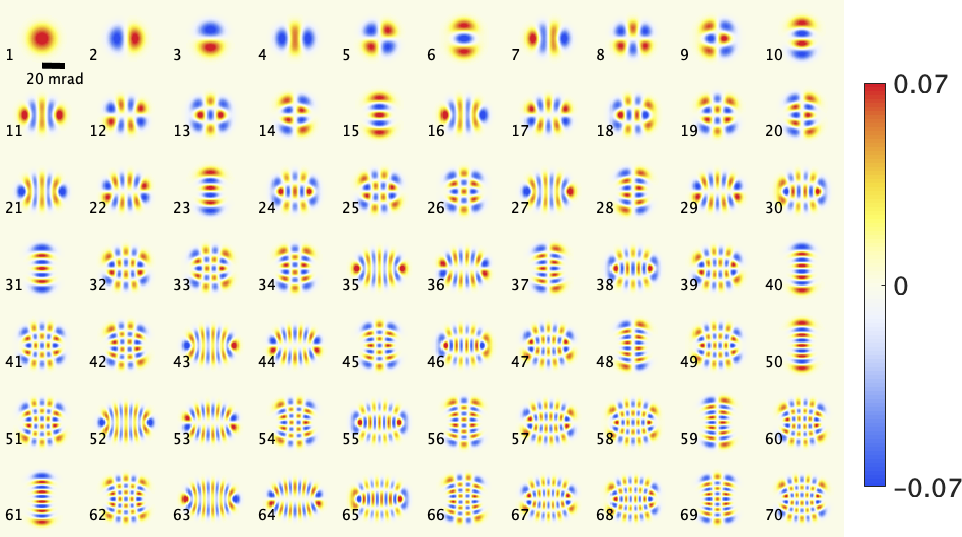}}
	\caption{First 70 simulated modes for PDC radiation. See Appendix~\ref{app:modes_theory} for details of the simulation.}
	\label{fig:modes_theory}
	\end{figure*}
	
Figures~\ref{fig:weights_exp},~\ref{fig:modes_exp} show the results of the eigenvalue decomposition of the re-organized $|G^{(1)}|$ with normalized weights. The distribution of the experimental weights shown in blue in Fig.~\ref{fig:weights_exp} follows an exponential decay, as expected from the theory~\cite{miatto_cartesian_2012}, and the effective number of coherent modes computed with Eq.~(\ref{eq:numbermodes}) is ${\cal K}_\text{ex} = 46\pm5$. The theoretical weights shown in red in Fig.~\ref{fig:weights_exp} show little discrepancy with respect to the one from the experiment.  
The calculated Schmidt number for the first two hundred simulated modes is ${\cal K}_\text{th} = 52$ which is close to the experimental value.
The eigenmodes reconstructed up to the 70th (Fig.~\ref{fig:modes_exp}) resemble the Hermite-Gauss modes, but show a small asymmetry that can be attributed to the ellipticity of the pump.

In App.~\ref{app:modes_theory} we present a theoretical model that allows to calculate numerically the first-order CF.
We find good agreement between the fit and the experiment
(in terms of angular width of the far-field PDC emission, angular width of Schmidt modes, distribution of weights) for the following values of the simulation parameters: phase mismatch parameter $\D_0 = -50\;\text{m}^{-1}$,
pump FWHM$_x =140\;\mu$m, pump ellipticity $\epsilon=1.2$, parametric gain $G=3.8$.
Obtained theoretical modes are shown in Fig.~\ref{fig:modes_theory}.
The small diagonal inclination of the experimental modes with respect to the simulated ones can be associated with the fact that in the experiment the transverse pump profile is elongated along the slightly rotated vertical axis.
Furthermore, the order of simulated modes (i.e. weights of the modes) is sensitive to the ellipticity parameter of the pump beam, as well as a phase mismatch, chosen in the simulation (see Appendix~\ref{app:modes_theory}).
This may explain that the order of the experimental and simulated modes is different in some cases, for example, for modes with numbers 15, 16, 17.
Also, a number of experimental mode pairs, e.g. $(10,11)$, $(20,21)$, $(29,30)$, resemble the hybridization of theoretical mode pairs $(10,11)$, $(20,21)$, $(28,30)$, respectively. It may be related to the proximity of the experimental mode weights (see Fig.~\ref{fig:weights_exp}) and degeneracy of the modes.

The agreement between the simulated and experimental modes can be tested with the fidelity, defined as
\begin{equation}
	{\cal F} = \int {d \rr\,{u_e}\left(\rr\right)u_s\left(\rr\right)},
	\label{eq:fidelity}
\end{equation}
where $u_{e}$ and $u_{s}$ correspond to normalized two-dimensional mode distributions in the experiment and in the simulation.
\begin{figure}[H]
	\center{\includegraphics[width=.8\linewidth]{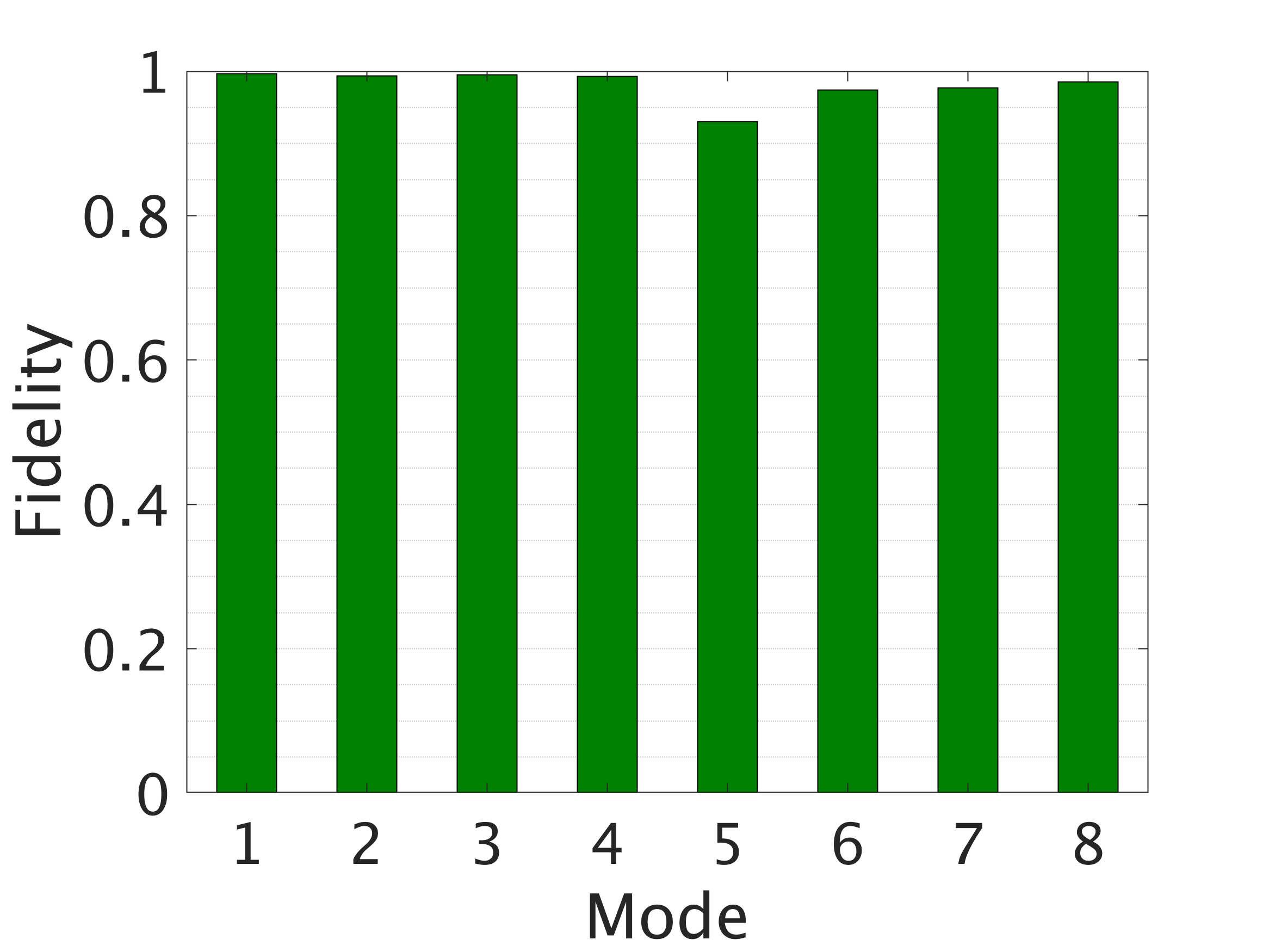}}
	\caption{Fidelity from Eq.~(\ref{eq:fidelity}) of the first eight reconstructed (see Fig.~\ref{fig:modes_exp}) and simulated (see Fig.~\ref{fig:modes_theory}) modes of the PDC radiation.}
	\label{fig:fidelity_PDC}
	\end{figure}
Fig.~\ref{fig:fidelity_PDC} shows very high fidelity for the first eight modes of the PDC source, confirming the accuracy of the reconstruction procedure. 
\begin{figure}[H]
	\center{\includegraphics[height=.47\linewidth]{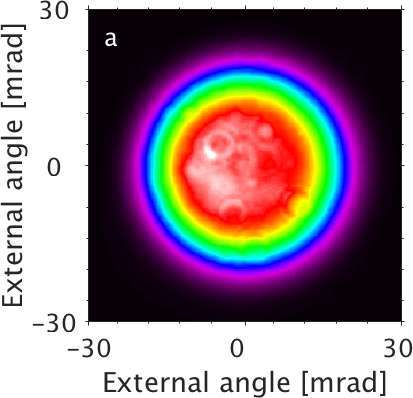}  \includegraphics[height=.47\linewidth]{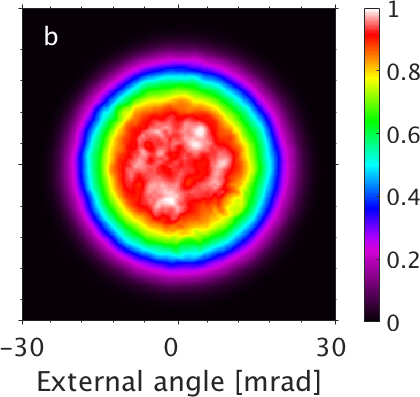}}
	\caption{(a) Average intensity spectrum from the experiment. (b) The intensity spectrum reconstructed from eigenmodes and eigenvalues.
	}
	\label{fig:meanspectrum}
	\end{figure}
	
Fig.~\ref{fig:meanspectrum} (a) shows the intensity spectrum of the far-field PDC emission averaged over the $3000$ spectra, with a FWHM of $\sim36$ mrad. The visible imperfections are due to the optics. To check the validity of our reconstruction method, we show in Fig.~\ref{fig:meanspectrum} (b) the intensity spectrum obtained from the modes and weights
\footnote{
The reconstructed intensity spectrum does not change whether we consider 70, 200 or all 10000 modes, because the weights decay exponentially.
}
by using Eq.~(\ref{eq:intensity}). The agreement between the two average spectra is good.

\subsection{Multimode fiber\label{sec:MMF}}

As a proof-of-principle experiment, we reconstruct the eigenmodes of a step-index fiber with $8.2$ $\mu$m core (SMF28 Thorlabs). Such a fiber supports a single mode in the infrared range, but it is multimode for visible light. The solutions to the Helmoltz equation for this weakly-guiding fiber are the well-known linearly polarized (LP) modes~\cite{saleh2007}. We compare the experimental results with the theory to validate our reconstruction method. In general, we point out that the coherent modes of a fiber coincide with the eigenmodes, hence the importance of this reconstruction method.

To generate light with a pseudo-thermal intensity distribution, we impinge the second-harmonic beam at 532 nm from the pulsed Nd:YAG laser described in Sec.~\ref{sec:highgainPDC} on a rotating ground glass disk (GGD), as shown in Fig.~\ref{fig:setup} (b). The speckle pattern obtained from the beam is coupled into the fiber with a $10$x microscope objective. The distances disk-objective and objective-fiber are chosen such that the disk is imaged onto the tip of the fiber with de-magnification $0.1$.
\begin{figure}[H]
	\center{\includegraphics[height=.47\linewidth]{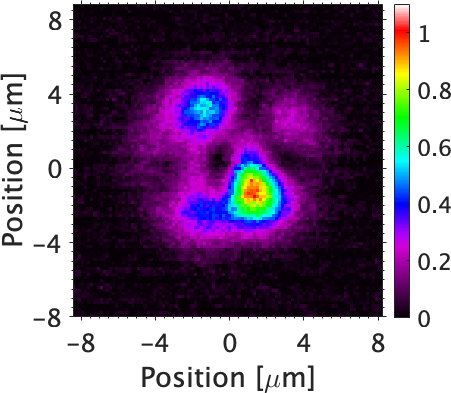}}
	\caption{Single-shot intensity distribution at the output of the fiber in the near field as a function of Cartesian coordinates.}
	\label{fig:shot_fiber}
	\end{figure}
	
The tip of the fiber at the output is then imaged on a CCD with another microscope objective providing a magnification of $71\pm1$. We acquire a set of $2000$ single-shot images, one of which is shown in Fig.~\ref{fig:shot_fiber} as a function of two Cartesian coordinates. Following the re-organisation procedure on the intensity distributions, we compute the covariance and by square root obtain the near-field first-order CF.
\begin{figure}[H]
	\center{\includegraphics[height=.465\linewidth]{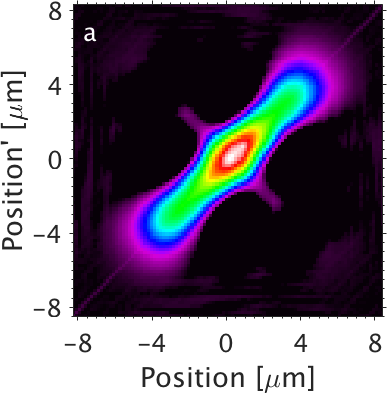}\includegraphics[height=.47\linewidth]{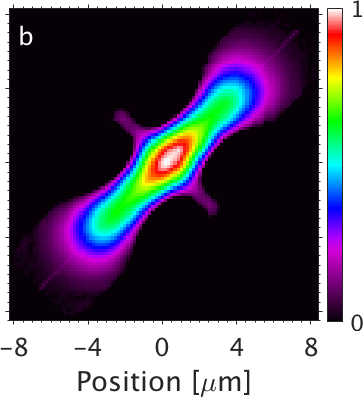}}
	\caption{$|G^{(1)}|$ distribution for the vertical cut of the output fiber intensity distribution at zero horizontal position} with (a) and without noise (b).
	\label{fig:covfiber}
	\end{figure}
	
In Fig.~\ref{fig:covfiber} (a), we only show $|G^{(1)}|$ distribution for the intensity profile at zero horizontal position. This distribution presents correlations not only along the main diagonal within a $2$ $\mu$m range, but also for the anti-diagonal (opposite positions close to the center of the core). The noise due to correlation with the camera dark noise is removed, as shown in Fig.~\ref{fig:covfiber} (b).

\begin{figure}[H]
	\center{\includegraphics[width=\linewidth]{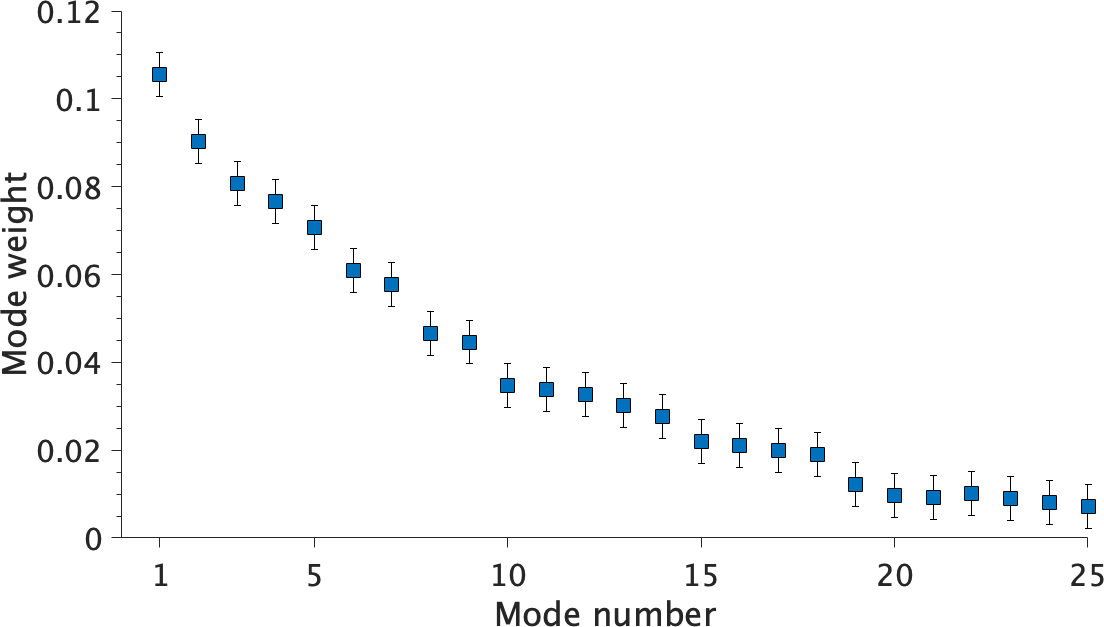}}
	\caption{Experimentally-reconstructed normalized weights of the fiber modes.
	}
	\label{fig:weights_exp_fiber}
	\end{figure}

\begin{figure*}[p]
	\center{\includegraphics[width=.85\linewidth]{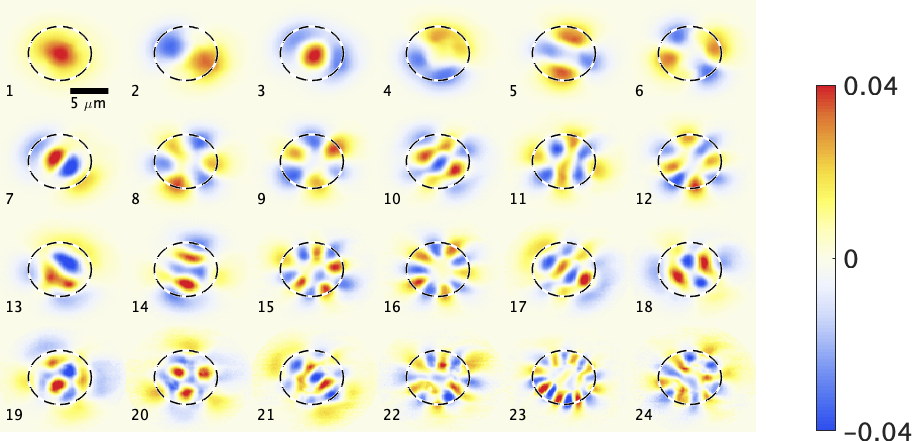}}
	\caption{First 24 reconstructed modes of the fiber. The black dashed line shows the core-cladding interface.}
	\label{fig:modes_exp_fiber}

	\center{\includegraphics[width=.92\linewidth]{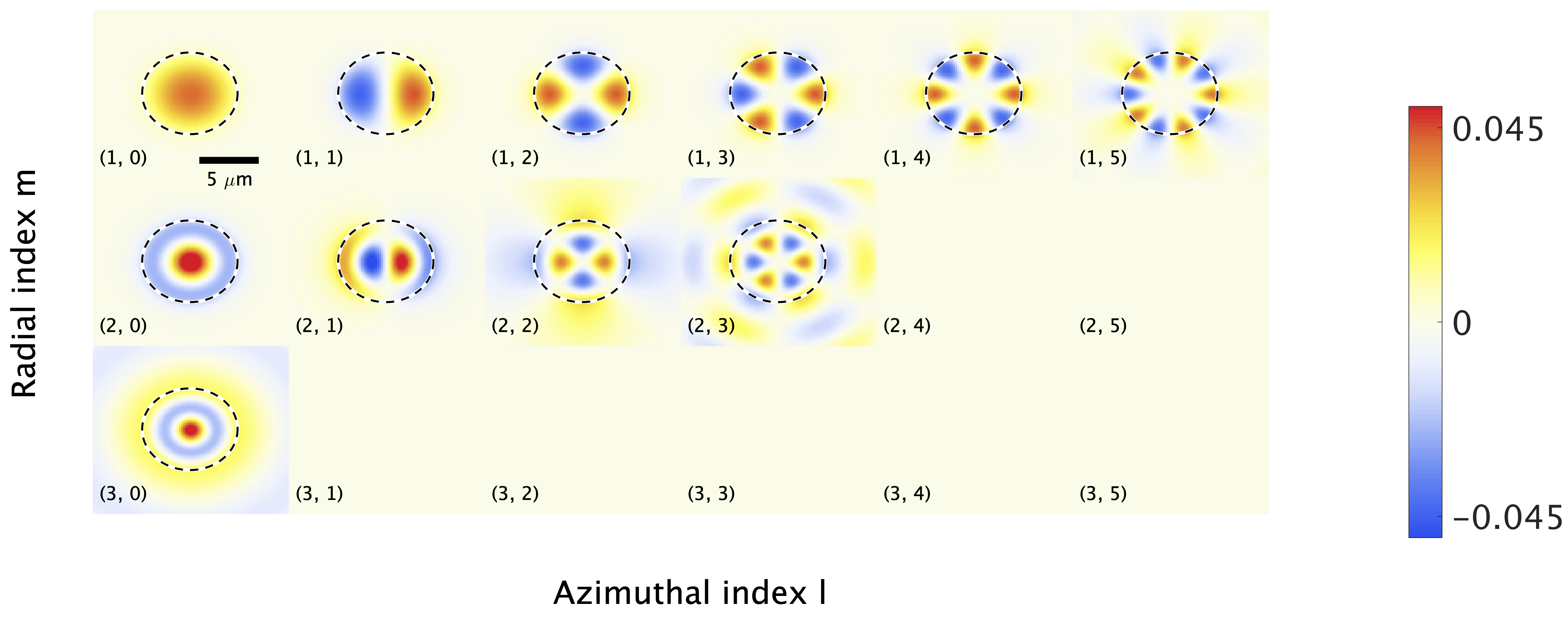}}
	\caption{LP modes, simulated for the fiber. The black dashed line represents the core-cladding interface.}
	\label{fig:modes_theory_fiber}
	\end{figure*}

We decompose the reorganised $|G^{(1)}|$ to obtain the eigenmodes and eigenvalues and the results are shown in the Figs.~\ref{fig:weights_exp_fiber},~\ref{fig:modes_exp_fiber}. 
The distribution of the weights has a power decay and the effective number of coherent modes computed from Eq.~(\ref{eq:numbermodes}) is $18\pm2$. This value is in good agreement with the theoretical value of 19 modes, obtained from the $V$ number of the fiber \cite{saleh2007}. The reconstructed eigenmodes resemble the LP modes, which are solutions of the Helmholtz equation for a multimode step-index fiber. The orientation in terms of azimuthal angle of the reconstructed modes may vary due to the fact that the fiber is not maintaining polarization. The simulated LP modes are shown in Fig.~\ref{fig:modes_theory_fiber} for comparison. Here, the modes are sorted according to the radial and azimuthal indices $(m, l)$ and the black dashed line represents the core-cladding interface. The blank slots are modes that cannot propagate inside the fiber. All the solutions with $l\neq0$ are related to a solution with azimuthal index $−l$. These modes are not shown because they differ only in the phase profile, but they should be counted to reach the number of supported modes.

\begin{figure}[H]
	\center{\includegraphics[width=.8\linewidth]{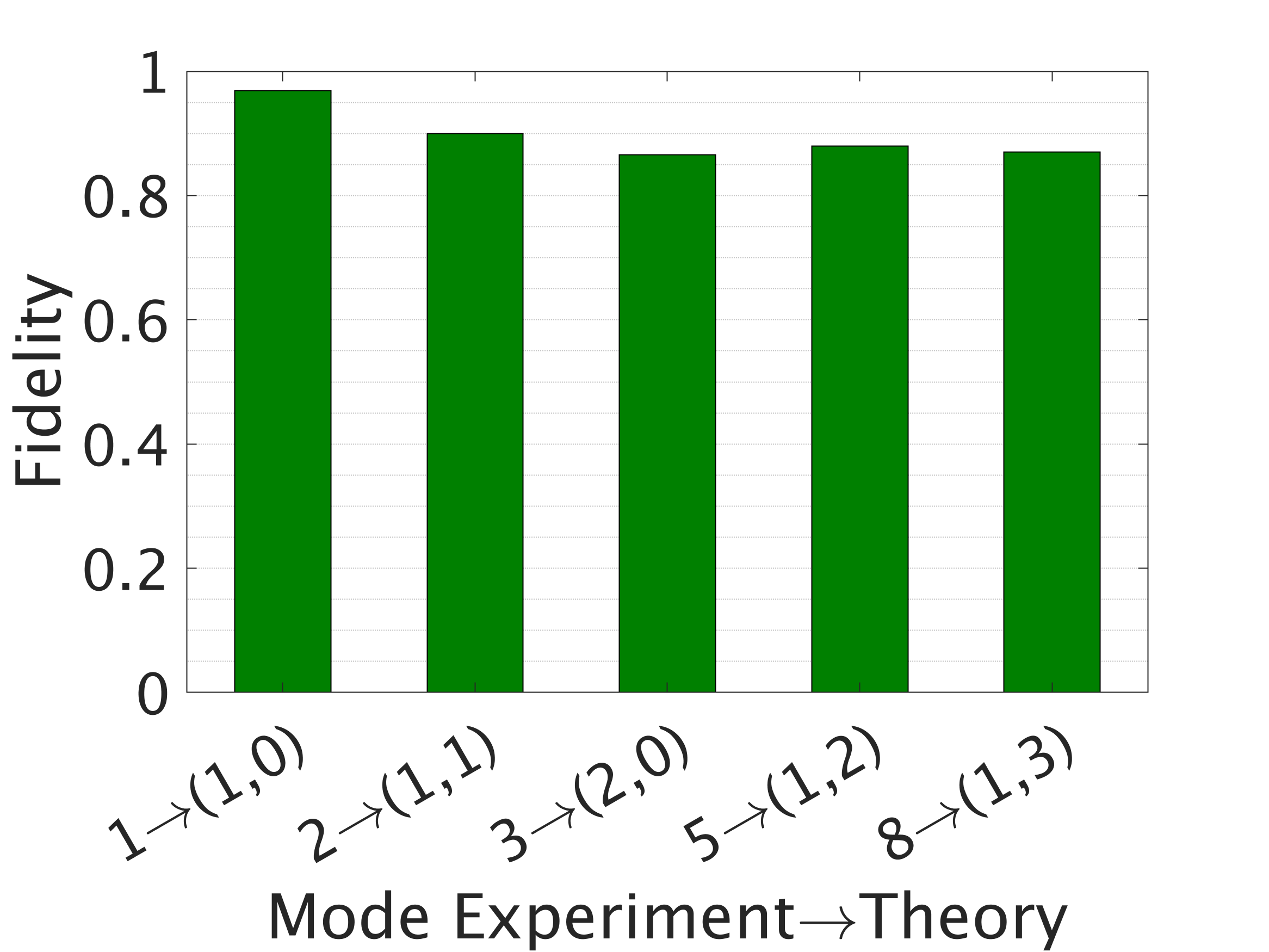}}
	\caption{Fidelity from Eq.~(\ref{eq:fidelity}) of five selected fiber modes from the reconstruction (see Fig.~\ref{fig:modes_exp_fiber}) and the theory (see Fig.~\ref{fig:modes_theory_fiber}).}
	\label{fig:fidelity_fiber}
	\end{figure}
	
To prove the validity of the reconstruction method, we evaluate the fidelity of the experimental and theoretical modes, defined in Eq.~(\ref{eq:fidelity}). Fig.~\ref{fig:fidelity_fiber} shows that the fidelity for the selected modes is always above $85\%$.

\begin{figure}[H]
	\center{\includegraphics[height=.5\linewidth]{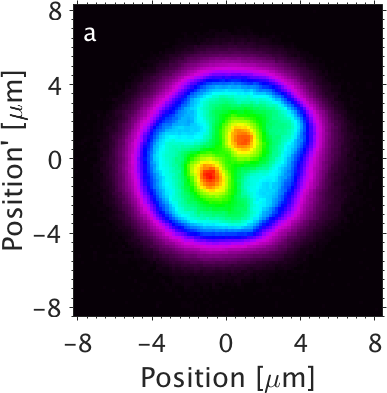}  \includegraphics[height=.5\linewidth]{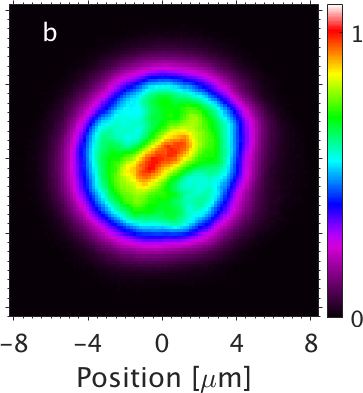}}
	\caption{(a) Average near-field intensity distribution from the experiment. (b) Intensity distribution reconstructed from the eigenmodes and eigenvalues.}
	\label{fig:meanspectrum_fiber}
	\end{figure}
	
The near-field intensity distribution averaged over the $2000$ spectra is shown in Fig.~\ref{fig:meanspectrum_fiber} (a). We point out that the mode content, and consequently the average intensity distribution, is highly dependent on the light coupling into the fiber; in our case, we adjusted the input objective to couple most light along the fiber axis. To check the validity of our reconstruction method, we show in Fig.~\ref{fig:meanspectrum_fiber} (b) the intensity distribution obtained from the modes and weights by using Eq.~(\ref{eq:intensity}). The agreement between the two distributions is good.

\section{Conclusion \& Outlook\label{sec:concl}}

We have presented an experimental method that allows to determine 2D profiles of the thermal field modes using direct measurement of the field intensity and offline analysis of its spatial fluctuations.
In particular, using this method, we solved two practical problems of classical and quantum optics: we correctly reconstructed two-dimensional modes of a multimode optical fiber and modes of down-converted radiation, in experimentally simple and efficient way.

The advantages of the proposed method are as follows.
First, modes are reconstructed based on a relatively simple measurement of field intensity correlations. It is similar to the advantages of the intensity interferometer in the HBT experiment compared to the field interferometer.
Second, the procedure allows one to reconstruct 2D profiles in a general case of spatially non-uniform and non-isotropic light fields.

The method is applicable to arbitrary thermal fields.
Furthermore, the analysis of multidimensional correlations proposed in this work can also be used for fields with non-thermal statistics. 
The limitation of the proposed procedure is that the correlation function of the field must be real and non-negative.

\acknowledgments
We thank Eugeny Mikhailov for drawing our attention to the procedure of reshaping of multidimensional arrays into arrays of lower dimension, which is a groundwork of this research,
and Ivan V. Sokolov for pointing out the connection of the proposed method to the HBT experiment. 
We wish to acknowledge Nicolas Y. Joly and Jonas Hammer for the fruitful discussions and the support received. 
VA thanks Cornelia Wild for support and acknowledges funding by Russian Science Foundation (project 17-19-01097-P). 
\clearpage
\newpage
\appendix

\section{Calculation of coherent modes of degenerate high-gain PDC}
\label{app:modes_theory}

Here we calculate coherent modes of the bright twin beams described in Sec.~\ref{sec:highgainPDC}.
We assume degenerate regime of PDC (wavelengths of signal and idler beams coincide) and use the following approximate expression for the signal/idler field correlation function \cite{Brambilla2004}
measured in the far field zone of the nonlinear crystal
	\begin{align}
	G^{(1)}(\qq, \qq') \propto &
	\iint\ud\rr e^{\ii (\qq-\qq')\rr}  A_\tp^2(\rr)\nn\\ 
	\times&\frac{\sinh\G(\qq,\rr)L}{\G(\qq, \rr)} \; 
	\frac{\sinh\G(\qq',\rr)L}{\G(\qq', \rr)}.
	\L{G1PDC}
	\end{align}
Here $\qq,\qq'$ are transverse wavevectors of the generated field. $A_\tp(\rr)$ is the spatial profile of the pump beam at the input face of the crystal and $\rr$ is the transverse coordinate in the cross-section of the pump beam. $L$ is the length of the nonlinear crystal. We use the following definitions:
	\begin{align}
	&\G(\qq, \rr) = \sqrt{\s^2 A_\tp^2(\rr) - \D^2(\qq)/4},\\
	&\Delta(\qq) = \Delta_0 - |\qq|^2/k_\ts.
	\end{align}
Here $\s$ is a constant proportional to the effective second-order susceptibility of the nonlinear crystal characterizing the downconversion process; $\Delta_0$ is a phase mismatch parameter that depends on the crystal dispersion; $k_{\tp, \ts} = 2\pi n_{\tp,\ts}/\lambda_{\tp,\ts}$ is the wavenumber of the pump/signal field at the central pump frequency/half of the pump frequency in the medium.

The expression (\ref{G1PDC}) is applicable for the case of the `narrow-band pump', i.e. the characteristic angular width of the pump beam is much smaller (but finite) than the angular width of PDC.
We assume that pump beam has flat phase front and the Gaussian transverse amplitude profile 
	\begin{align}
	& A_\tp(x,y) = \frac{G}{\s L}\; \text{exp}\left(-(x^2+y^2/\epsilon^2)/2w_\tp^2\right)
	\L{Gaussian_pump_profile}
	\end{align}
Here $G$ is equal to the parametric gain \cite{sharapova_properties_2020}. We also introduce the pump ellipticity parameter $\epsilon$ to take into account the possible asymmetry of the pump beam: the x-axis beam size is characterized by a full-width at half-maximum $\text{FWHM}_x =  2\sqrt{\ln{2}}\;w_\tp$ (with $w_\tp$ being the beam waist) and the y-axis size is $\text{FWHM}_y~=~\epsilon~\;~\text{FWHM}_x$.
The characteristic angular width of the pump beam defined as in Ref.~\cite{Brambilla2004} is $\dd q_0 = \sqrt{2}/w_p$, while for PDC it is $q_0 = \sqrt{k_\tp/2L}$ assuming wavelength degeneracy.
For the experimental conditions of $L=2$mm, $\l_\tp=355$ nm and $n_\tp=1.7$, the `narrow-band pump' condition is fulfilled when FWHM $\gg30\mu$m.

In the expression (\ref{G1PDC}) we also assume that the pump beam i) does not diffract
\footnote{For a pump with FWHM$=100\mu$m, the Rayleigh length is 10cm and substantially exceeds the experimental crystal length.}, ii) has no walk-off in the crystal, iii) is monochromatic.

Due to imperfect crystal alignment, the phase mismatch parameter $\Delta_0$ can be slightly non-zero. Such deviation is hard to fix in the experiment but affects the shape of the intensity distribution of signal/idler photons.

Below we present the calculation for the following parameter values: phase mismatch parameter $\D_0 = -50\;\text{m}^{-1}$,
pump FWHM$_x =140\;\mu$m, ellipticity $\epsilon=1.2$, parametric gain $G=3.8$.

\begin{figure}[H]
	\center{\includegraphics[height=.7\linewidth]{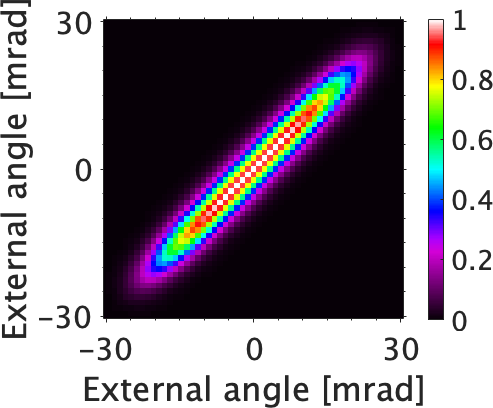}}
	\caption{
	Correlation function for the PDC emission (\ref{G1PDC}) with the vertical coordinate fixed to zero (i.e. $q_y = q_y'=0$) as a function of the external angles, defined as $\theta = q_x/k_\ts$ and $\theta' = q_x'/k_\ts$.
	}
	\label{fig:G1_th_high}
	\end{figure}

Fig.~\ref{fig:G1_th_high} shows calculated correlation function for the PDC emission (\ref{G1PDC}) with vertical coordinate fixed to zero (i.e. $q_y = q_y'=0$) as a function of the external angles, defined as $\theta = q_x/k_\ts$ and $\theta' = q_x'/k_\ts$.

\begin{figure}[H]
	\center{\includegraphics[height=.7\linewidth]{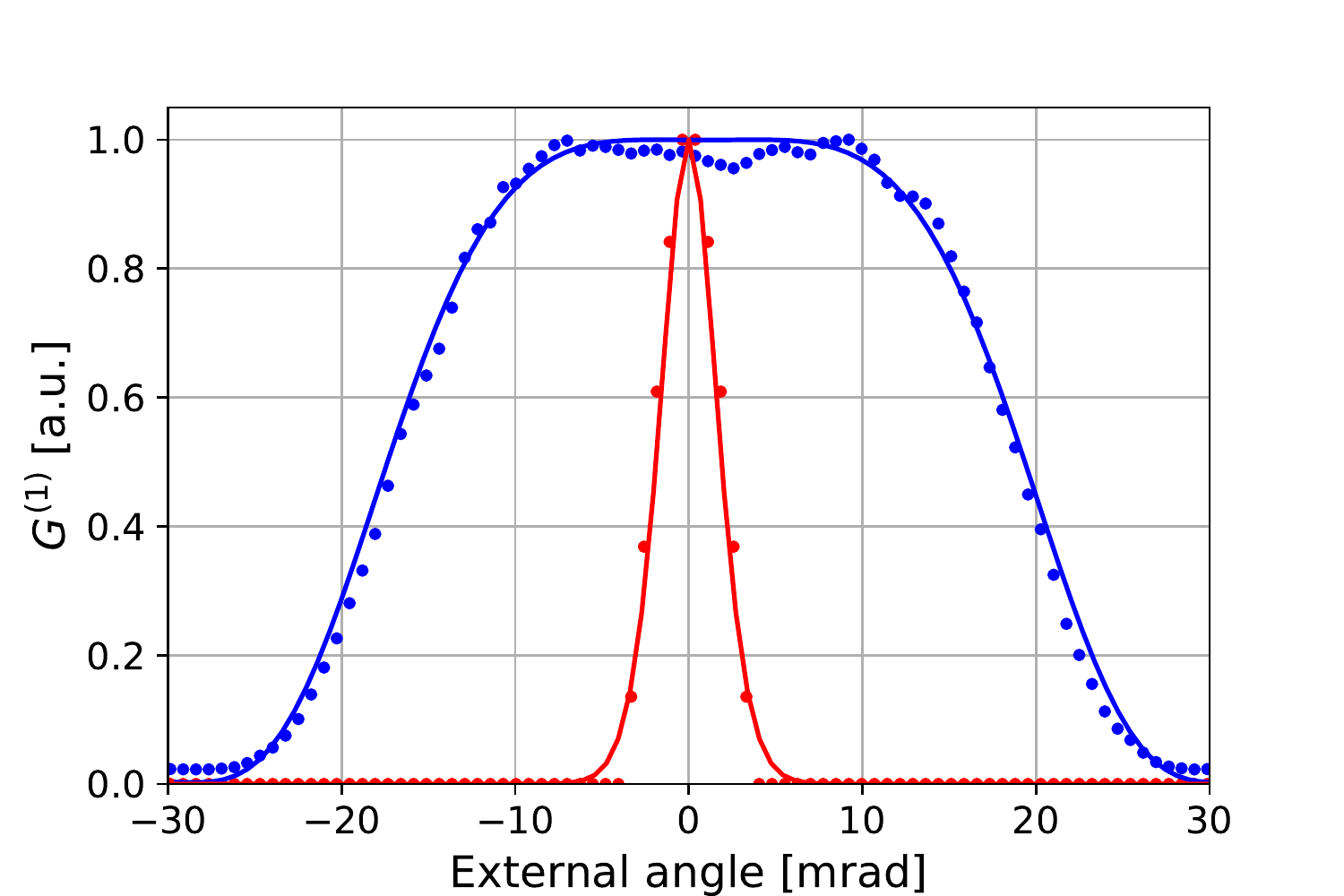}}
	\caption{Diagonal (blue) and anti-diagonal (red) values of the experimental (dots) and simulated (solid) correlation functions shown in Fig.~\ref{fig:covPDC}b and Fig.~\ref{fig:G1_th_high}, respectively.
	}
	\label{fig:covPDC_th_VS_exp}
	\end{figure}

Fig.~\ref{fig:covPDC_th_VS_exp} presents comparison of the diagonal/anti-diagonal values of the simulated (Fig.~\ref{fig:G1_th_high}) and experimental (Fig.~\ref{fig:covPDC} b) correlation functions.
It is worth mentioning that diagonal values of the correlation function determine the intensity profile of the signal beam (in this case, at $q_y=0$), according to Eq.~(\ref{I_def}).

\begin{figure}[H]
	\center{\includegraphics[width=.7\linewidth]{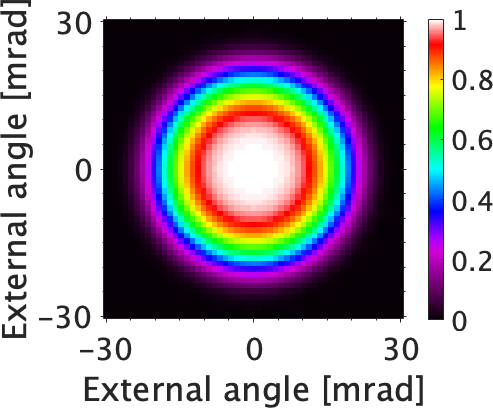}}
	\caption{Simulated average intensity.
	}
	\label{fig:intensity_th_high}
	\end{figure}

Fig.~\ref{fig:intensity_th_high} shows the calculated transverse signal intensity distribution.

\begin{figure}[H]
	\center{\includegraphics[width=.7\linewidth]{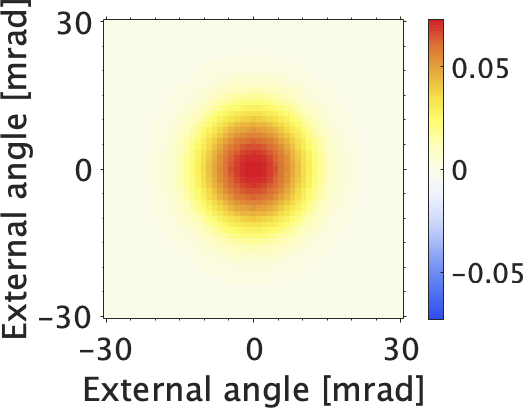}}
	\caption{Spatial profile of the first calculated Schmidt mode.
	}
	\label{fig:first-mode_high}
	\end{figure}

Further, we reorganize the values of the calculated correlation function into a 2-D array following the procedure shown in the expression (\ref{G1_2D}) and we diagonalize the array using Eq.~(\ref{eigen_problem}).
The eigenvectors obtained with diagonalization of the correlation function are transformed into matrices according to the procedure presented in (\ref{vector-to-matrix}). As a result one gets two-dimensional Schmidt mode profiles which are shown in Fig.~\ref{fig:modes_theory}.
Particularly, Fig.~\ref{fig:first-mode_high} shows the spatial profile of the first simulated Schmidt mode.

\bibliography{2Dmodes}

\end{document}